\documentclass[11pt]{article}
\usepackage{graphicx}
\usepackage[margin=1.25in]{geometry}
\usepackage[usenames,dvipsnames,table]{xcolor}
\usepackage{url}
\usepackage{caption}
\usepackage{amssymb,amsmath}
\usepackage[colorlinks = true,
            linkcolor = blue,
            urlcolor  = blue,
            citecolor = blue,
            anchorcolor = blue]{hyperref}

\usepackage[utf8]{inputenc}
\usepackage{cite}
\usepackage{tabularx}

\newcommand\pubnumber{KEK-TH-2400}
\newcommand\pubdate{\relax}

\def\Title#1{\begin{center} {\LARGE #1 } \end{center}}
\def\Author#1{\begin{center}{ \sc #1} \end{center}}
\def\Address#1{\begin{center}\small{ \it #1} \end{center}}

\newcommand\pubblock{\rightline{\begin{tabular}{l} \pubnumber\\
         \pubdate \end{tabular}}}
\newenvironment{Abstract}%
  {\begin{center}\begin{minipage}{.88\textwidth}\hfil ABSTRACT\hfil\par\medskip}%
  {\end{minipage}\end{center}}

\newcommand{\bhline}[1]{\noalign{\hrule height #1}}

\def\ie{{\it i.e.}}
\def\eg{{\it e.g.}}

\def\beq{\begin{equation}}
\def\eeq#1{\label{#1}\end{equation}}
\def\eeqn{\end{equation}}

\newenvironment{Eqnarray}%
   {\arraycolsep 0.14em\begin{eqnarray}}{\end{eqnarray}}
\def\beqa{\begin{Eqnarray}}
\def\eeqa#1{\label{#1}\end{Eqnarray}}
\def\eeqan{\end{Eqnarray}}

\let\bar=\overbar

\def\lsim{\mathrel{\raise.3ex\hbox{$<$\kern-.75em\lower1ex\hbox{$\sim$}}}}
\def\gsim{\mathrel{\raise.3ex\hbox{$>$\kern-.75em\lower1ex\hbox{$\sim$}}}}

\def\del{\partial}
\def\Dslash{\not{\hbox{\kern-4pt $D$}}}
\def\dslash{\not{\hbox{\kern-2pt $\del$}}}
\def\pslash{\not{\hbox{\kern-2pt $p$}}}
\def\ETmiss{\not{\hbox{\kern-4pt $E$}}_T}

\def\Dlr{\mathrel{\raise1.5ex\hbox{$\leftrightarrow$\kern-1em\lower1.5ex\hbox{$D$}}}}

\def\MSB{{\bar{M \kern -2pt S}}}
\def\msb{{\bar{\scriptsize M \kern -1pt S}}}

\def\drb{{\bar{\scriptsize D \kern -1pt R}}}

\mathchardef\mhyphen="2D

\newcommand\snowmass{\begin{center}\rule[-0.2in]{\hsize}{0.01in}\\\rule{\hsize}{0.01in}\\
\vskip 0.1in Submitted to the  Proceedings of the US Community Study\\
on the Future of Particle Physics (Snowmass 2021)\\
\rule{\hsize}{0.01in}\\\rule[+0.2in]{\hsize}{0.01in} \end{center}}

\newcommand\w[1]{_{\mathrm{#1}}}
\newcommand\unit[1]{\,\mathrm{#1}}

\newcommand\GeV{\unit{GeV}}
\newcommand\TeV{\unit{TeV}}

\newcommand\iab{\unit{ab^{-1}}}

\newcommand\Damu{\Delta a_\mu}
\newcommand\amu[1][\relax]{\ifx#1\relax{a_\mu}\else{a_\mu^{\mathrm{#1}}}\fi}
\newcommand\amubino{a_\mu^{(\tilde B)}}

\newcommand\mmuLR{m_{\tilde\mu\mathrm{LR}}}
\newcommand\mlLR{m_{\tilde l\mathrm{LR}}}
\newcommand\mtauLR{m_{\tilde\tau\mathrm{LR}}}
\begin{document}

\pubblock

\Title{\bf{
Stau study at the ILC and its implication \\[0.25em]
for the muon \boldmath{$g-2$} anomaly
}
}

\bigskip

\Author{
  Motoi Endo$^{(a,b,c)}$,
  Koichi Hamaguchi$^{(c,d)}$,
  Sho Iwamoto$^{(e)}$,
  Shin-ichi Kawada$^{(f)}$,
  Teppei Kitahara$^{(g,h)}$,
  Takeo Moroi$^{(c,d)}$
  and
  Taikan Suehara$^{(i)}$
}

\medskip

\Address{
$^{(a)}$ KEK Theory Center, IPNS, KEK, Tsukuba, Ibaraki 305--0801, Japan
\\
$^{(b)}$ The Graduate University of Advanced Studies (Sokendai), Tsukuba, Ibaraki 305--0801, Japan
\\
$^{(c)}$ Kavli IPMU (WPI), UTIAS, The University of Tokyo, Kashiwa, Chiba 277--8583, Japan
\\
$^{(d)}$ Department of Physics, The University of Tokyo, Bunkyo-ku, Tokyo 113--0033, Japan
\\
$^{(e)}$ ELTE E\a"otv\a"os Lor\a'and University, P\a'azm\a'any P\a'eter s\a'et\a'any 1/A, Budapest H-1117, Hungary
\\
$^{(f)}$ IPNS, KEK, Tsukuba, Ibaraki 305--0801, Japan
\\
$^{(g)}$ Institute for Advanced Research, Nagoya University, Nagoya 464--8601, Japan
\\
$^{(h)}$ Kobayashi-Maskawa Institute for the Origin of Particles and the Universe, Nagoya University,  Nagoya 464--8602, Japan
\\
$^{(i)}$ Department of Physics, Kyushu University, Fukuoka 819--0395, Japan
}

\medskip

 \begin{Abstract}
\noindent
Once all the sleptons as well as the Bino are observed at the ILC, the
Bino contribution to the muon anomalous magnetic dipole moment (muon
$g-2$) in supersymmetric (SUSY) models can be reconstructed.
Motivated by the recently confirmed muon $g-2$ anomaly, we examine the
reconstruction accuracy at the ILC with $\sqrt{s}=500\,$GeV. For this
purpose, measurements of stau parameters are important.  We
quantitatively study the determination of the mass and mixing
parameters of the staus at the ILC.  Furthermore, we discuss the
implication of the stau study to the reconstruction of the SUSY
contribution to the muon $g-2$.  At the benchmark point of our choice, 
  we find that the SUSY contribution to the muon $g-2$ can be determined with a precision of $\sim  1\%$ at the ILC.
\end{Abstract}

\snowmass
\begin{NoHyper}
  \renewcommand\thefootnote{}
  \footnotetext{This is a preliminary study performed in the framework of the ILD concept group.}
\end{NoHyper}

\renewcommand{\thefootnote}{\#\arabic{footnote}}
\setcounter{footnote}{0}

\section{Introduction}

The Fermilab experiment of measuring the muon anomalous magnetic moment (muon $g-2$) confirmed the long-standing discrepancy between its measured value at the Brookhaven experiment~\cite{Bennett:2002jb,Bennett:2004pv,Bennett:2006fi,Abi:2021gix} and the Standard Model (SM) prediction~\cite{Aoyama:2020ynm},
\begin{align}
 \amu[\text{BNL}+\text{FNAL}] &=
\left(11\,659\,206.1 \pm 4.1 \right) \times 10^{-10}\,,
\\
 \amu[\text{SM}] &= \left( 11\,659\,181.0 \pm 4.3 \right) \times 10^{-10}\,,
\end{align}
which amounts to a discrepancy at the $4.2\sigma$ level:
\begin{equation}
 \Delta a_{\mu}
 \equiv  \amu[\text{BNL}+\text{FNAL}]  - \amu[\text{SM}]
 =      \left( 25.1\pm 5.9\right) \times 10^{-10}\,,
 \label{eq:Deltaamu}
\end{equation}
where $a_\mu\equiv (g_\mu-2)/2$ with $g_\mu$ being the muon magnetic moment.\footnote{%
  This SM value is based on the data-driven method for determination of the hadronic vacuum-polarization contribution.
  On the other hand,
  a lattice method provides a different estimation,
  with which the SM prediction is consistent with the measured muon $g-2$, while producing a new tension.
  See, \eg, footnotes $\#1$--3 of Ref.~\cite{Endo:2021zal} for further details of the SM prediction.
}
This discrepancy may be a hint for physics beyond the SM.
Noting that $\Delta a_\mu$ is as large as the SM electroweak contribution to $\amu$, new particles with a mass at the electroweak scale may be the source of this discrepancy.

Low-energy supersymmetry (SUSY) is one of such solutions~\cite{Lopez:1993vi,Chattopadhyay:1995ae,Moroi:1995yh}.
It predicts new particles (SUSY particles) that
interact with muons and photons, yielding extra contribution to $a_\mu$, which we call $\amu[SUSY]$.
If their masses are at the sub-TeV scale, $\amu[SUSY]$ can be sizable enough to solve the discrepancy.
This feature does not harm other benefits of SUSY.
In particular, with the thermal freeze-out mechanism, the lightest SUSY particle (LSP) can account for dark matter, with a relic density consistent with the observed value.
For example, Ref.~\cite{Endo:2021zal} pointed out that there exist parameter spaces of the minimal supersymmetric Standard Model (MSSM) in which the muon $g-2$ anomaly is well explained, the LSP becomes the dark matter candidate with the observed relic density, and the latest constraints from collider experiments and dark matter direct detections are avoided.

As the SUSY solution to the muon $g-2$ anomaly requires some SUSY particles to be relatively light, those particles are important targets of future collider experiments.
More interestingly, by determining the properties of the SUSY particles, the SUSY contribution $\amu[SUSY]$ can be reconstructed to confirm that the anomaly truly originates in SUSY particles~\cite{Endo:2013xka}.
For this purpose, masses and coupling constants of the SUSY particles should be precisely measured.

\begin{figure}[t]
  \centering
  \includegraphics[width=0.3\linewidth]{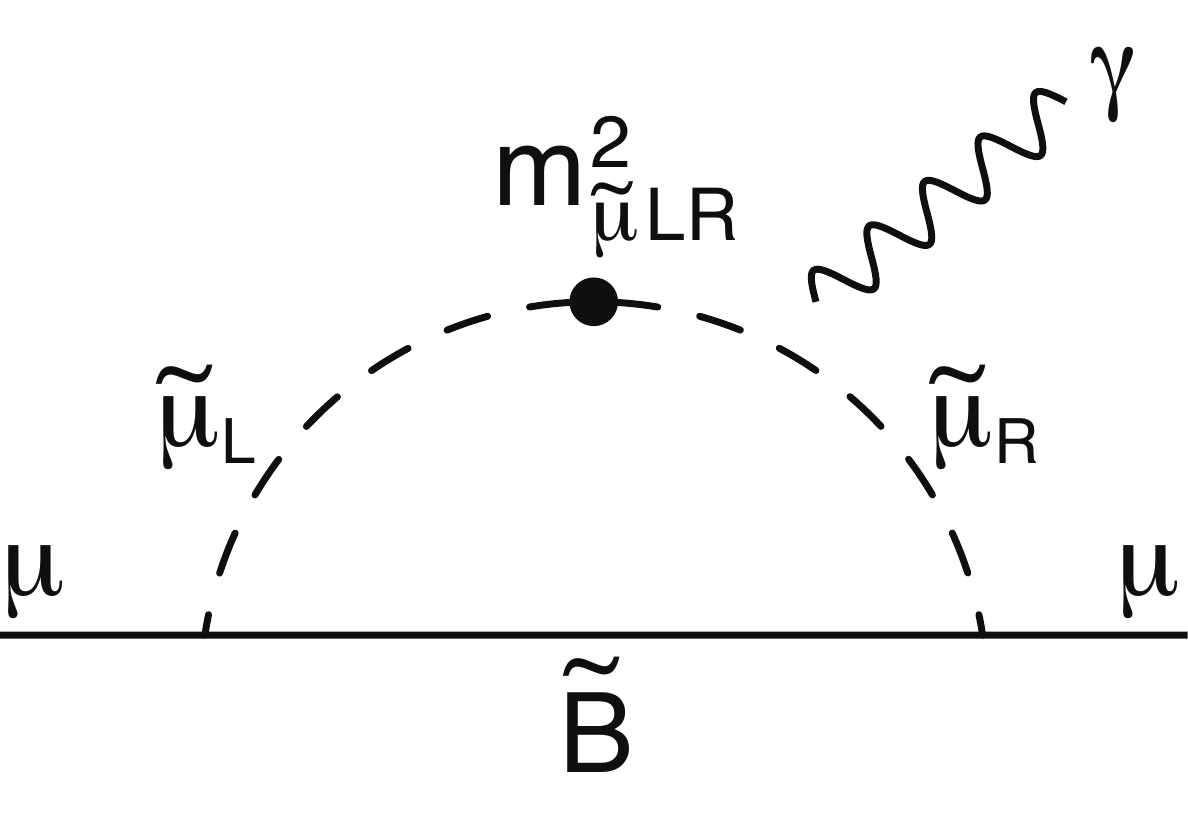}
  \caption{The Bino-smuon loop diagram contributing to $\amu[SUSY]$.
}
  \label{fig:binosmu}
\end{figure}

The International Linear Collider (ILC) is an ideal facility to perform such measurements.
Although the SUSY contribution $\amu[SUSY]$ comes from various diagrams, we focus on the so-called Bino-smuon diagram (Fig.~\ref{fig:binosmu});
we denote its contribution by $\amubino$.
In the parameter space with a large Higgsino mass parameter $\mu\gg1\TeV$, it tends to dominate the SUSY contribution $\amu[SUSY]$.
The four benchmark points in Ref.~\cite{Endo:2021zal}, BLR1--4, are such examples;
Binos $\tilde B$ and smuons $\tilde\mu\w{L,R}$ are as light as 100--$200\GeV$ and $\mu\gg1\TeV$, which realizes $\amu[SUSY]\simeq\amubino\simeq\Damu$.
In addition, staus $\tilde\tau\w{1,2}$ and tau-sneutrinos $\tilde\nu_{\tau}$ have masses smaller  than $250\GeV$ and coannihilation yields the correct dark matter relic abundance.
These particles are thus within the reach of the ILC with center-of-mass energy $\sqrt{s}=500\GeV$, and
we can reconstruct $\amubino$ by measuring their properties.

For the reconstruction of $a_\mu^{(\tilde{B})}$, it is necessary to know (i) masses of smuons, (ii) Bino (\ie, the lightest neutralino) mass, (iii) lepton-slepton-Bino couplings, and (iv) left-right mixing of the smuons $\mmuLR^2$ (cf.\ Eq.~\eqref{eq:massmatrix}) as depicted in  Fig.~\ref{fig:binosmu}.
Among them, (i)--(iii) can be precisely determined by using the smuon and selectron production processes, as summarized in Ref.~\cite{Endo:2013xka}.
On the contrary, the determination of the left-right mixing of the smuons is highly non-trivial.
Information about the smuon left-right mixing is difficult to obtain from the smuon production processes at the ILC because $\mmuLR^2$ is proportional to the muon mass and is tiny relative to the squared smuon masses.

In Ref.~\cite{Endo:2013xka}, it has been proposed to rely on a SUSY
relation to determine $\mmuLR^2$.  In the parameter region of our
interest, \ie, if the Higgsino mass $\mu$ is much larger than the
slepton trilinear couplings (the $A_l$ parameters), the left-right mixings of
the sleptons are approximated by $\mlLR^2\simeq -m_l\mu\tan\beta$ with
$l=e$, $\mu$, and $\tau$ and $\tan\beta$ being
the ratio of the vacuum expectation values of up- and
down-type Higgs bosons.\footnote{%
This relation holds in the limit of
  $|A_{l}|\ll |\mu| \tan\beta$.
}
Then, $\mmuLR^2$ can be related to $\mtauLR^2$ as
\begin{equation}
  m_{\tilde{\mu}{LR}}^2
  =
  \frac{m_\mu}{m_\tau} m_{\tilde{\tau}{LR}}^2\,.
  \label{eq:mlrratio}
\end{equation}
The left-right mixing of the staus is an order of magnitude larger
than that of the smuons and hence is easier to measure at the ILC.
This provides a strong motivation to study stau properties at the ILC
in connection with the muon $g-2$ anomaly.
The ILC's potential to determine the stau properties in the parameter
region motivated by the muon $g-2$ anomaly, however,
is not yet well understood.

Motivated by the possibility to reconstruct the SUSY contribution to the muon $g-2$, we investigate the prospect of measuring stau properties at the ILC.
We pay particular attention to the parameter region suggested by the muon $g-2$ anomaly.  We perform a detailed Monte Carlo (MC) analysis to see the accuracy of the ILC measurements of the stau mass and mixing parameters.
We also qualitatively discuss the implication of the stau study for the reconstruction of $\amu[SUSY]$.

This work is organized as follows. In Section \ref{sec:study}, we explain our basic strategy to study the stau properties at the ILC.  In Section \ref{sec:stau}, we show the results of our MC analysis.  Implication of the stau study to the muon $g-2$ anomaly is discussed in Section \ref{sec:amu}.  The results are summarized in Section \ref{sec:summary}.

\section{Stau Study at the ILC: Basic Strategy}\label{sec:study}

Let us begin with our strategy for the determination of the stau property at the ILC, in particular the stau left-right parameter $\mtauLR^2$, which is used in the reconstruction of $\amu[(\tilde B)]$ in Sec.~\ref{sec:amu}.

The stau mass eigenstates are denoted as $\tilde{\tau}_1$ and $\tilde{\tau}_2$
(with their masses $m_{\tilde{\tau}_1}$ and $m_{\tilde{\tau}_2}$),
while the gauge eigenstates as $\tilde{\tau}_L$ and $\tilde{\tau}_R$
(where $\tilde{\tau}_L$ is embedded in $SU(2)_L$ doublet while
$\tilde{\tau}_R$ is $SU(2)_L$ singlet).  The mass matrix in the gauge
eigenbasis is denoted as
\begin{align}
  \mathcal{L} \supset
  -   \begin{pmatrix}
  \tilde{\tau}_L^\ast &
  \tilde{\tau}_R^\ast
  \end{pmatrix}\mathcal{M}_{\tilde{\tau}}^2
  \begin{pmatrix}
  \tilde{\tau}_L\\
  \tilde{\tau}_R
  \end{pmatrix}
\quad \text{with} \quad \mathcal{M}_{\tilde{\tau}}^2 =  \begin{pmatrix}
    m_{\tilde{\tau}LL}^2 & m_{\tilde\tau{LR}}^2
    \\
    m_{\tilde\tau{LR}}^{2} &
    m_{\tilde{\tau}RR}^2
    \end{pmatrix}\,.
    \label{eq:massmatrix}
\end{align}
Here, we assume that CP violation in the MSSM sector is negligible so that
$m_{\tilde\tau{LR}}^{2}$ can be taken to be real.  With diagonalizing the stau mass matrix, the mass and gauge eigenstates are related by using the stau mixing angle $\theta_{\tilde\tau}$ as
\begin{equation}
  \begin{pmatrix}
  \tilde{\tau}_1\\
  \tilde{\tau}_2
  \end{pmatrix}
  =
  \begin{pmatrix}
  \cos\theta_{\tilde\tau} & \sin\theta_{\tilde\tau} \\
 -\sin\theta_{\tilde\tau} & \cos\theta_{\tilde\tau}
  \end{pmatrix}
  \begin{pmatrix}
  \tilde{\tau}_L\\
  \tilde{\tau}_R
  \end{pmatrix}
  \equiv
  U_{\tilde{\tau}}
  \begin{pmatrix}
  \tilde{\tau}_L\\
  \tilde{\tau}_R
  \end{pmatrix}\,,
\end{equation}
where $m_{\tilde{\tau}_1}< m_{\tilde{\tau}_2}$ and $0 \leq
\theta_{\tilde{\tau}}< \pi$.

In the smuon sector, we obtain similar
relations in relating the mass and gauge eigenstates; the above
argument is applicable to the smuons with replacing
$\tilde{\tau}\rightarrow\tilde{\mu}$. (In the smuon sector, we also
take $m_{\tilde{\mu}_1}< m_{\tilde{\mu}_2}$ and $0 \leq
\theta_{\tilde{\mu}}< \pi$.)

Our primary purpose is to determine $m_{\tilde\tau{LR}}^{2}$, which is related to the mass and mixing parameters of the staus as
\begin{equation}
  m_{\tilde\tau{LR}}^2 =
  \frac{1}{2}
  \left(m_{\tilde\tau_1}^2 - m_{\tilde\tau_2}^2\right)
  \sin 2 \theta_{\tilde\tau} \,.
  \label{eq:sin2thl}
\end{equation}
For this purpose, we use the stau pair-production processes
\begin{align}
  e^+ e^- \rightarrow \tilde{\tau}_i^* \tilde{\tau}_j~~~
  (i,j=1,2)\,.
\end{align}
Their cross sections are given by \cite{Nojiri:1994it,Boos:2003vf}
\begin{equation}
\begin{split}
\sigma(e^+e^-\to\tilde\tau_i^\ast\tilde\tau_j) &=
\frac{8\pi\alpha^2}{3s} \beta\w{f}^3 \bigg[
c_{ij}^2 \frac{\Delta_Z^2}{\sin^42\theta_W}(\mathcal{P}_{-+}L^2 + \mathcal{P}_{+-}R^2)
\\&\qquad
+\delta_{ij} \frac{1}{16} (\mathcal{P}_{-+} + \mathcal{P}_{+-})
+\delta_{ij}c_{ij}\frac{\Delta_Z}{2\sin^22\theta_W}(\mathcal{P}_{-+}L + \mathcal{P}_{+-}R)
\bigg]\,,
\end{split}
\label{eq:CrossSection}
\end{equation}
where
\begin{align}
&\beta\w{f}^2 = \left[ 1-\frac{(m_{\tilde\tau_i}+m_{\tilde\tau_j})^2}{s}\right]\left[1-\frac{(m_{\tilde\tau_i}-m_{\tilde\tau_j})^2}{s}\right]\,, \\
&\Delta_Z = \frac{s}{s-m_Z^2}\,, \\
&c_{11/22} = \frac{1}{2} \left[L+R \pm (L-R) \cos 2\theta_{\tilde{\tau}} \right]\,,\\
&c_{12} = c_{21} = \frac{1}{2}(L-R)\sin 2\theta_{\tilde{\tau}}\,,\\
&L=-\frac{1}{2}+\sin^2\theta_W\,,\\
&R=\sin^2\theta_W\,.
\end{align}
The beam polarizations are parameterized as $\mathcal{P}_{\mp\pm} = (1
\mp P_{e^-}) (1 \pm P_{e^+})$, where $P_{e}=\pm1$
corresponds to particles fully-polarized with the helicity $\pm1$.
As one can see, the cross sections are
sensitive to {the beam polarizations and} the stau mixing angle
$\theta_{\tilde{\tau}}$ as well as the stau masses.
Physics of the stau sector is closely discussed in Refs.~\cite{Nojiri:1996fp,Bechtle:2009em,Berggren:2015qua}.

Other important observables are the endpoints of the energy
  distributions of the decay products of staus.  At the benchmark
  point of our choice, the staus decay as
\begin{align}
  \tilde{\tau}_i \rightarrow \tau \tilde{B}
\end{align}
and the tau decays leptonically or hadronically.  At the ILC, we can
identify the visible decay products of $\tau$ and measure their
energy.  Considering the stau with its energy $E_{\tilde{\tau}_i}$,
the energy of the visible decay products, $E_{\tau}$, is bounded from
above as
\begin{align}
  E_{\tau} < E_{+} (\tilde{\tau}_i) \equiv
  \frac{m_{\tilde{\tau}_i}^2-m_{\tilde{\chi}^0_1}^2}{2m_{\tilde{\tau}_i}^2}
  \left(
    E_{\tilde{\tau}_i} + \sqrt{E_{\tilde{\tau}_i}^2-m_{\tilde{\tau}_i}^2}
    \right)\,,
    \label{eq:endpoint}
\end{align}
where $m_{\tilde{\chi}^0_1}$ is the lightest neutralino mass. (In discussing the endpoints, the
$\tau$ lepton mass can be ignored with good accuracy.)  The
endpoints of the energy distributions of the decay products of the staus
provide information about the stau masses.

In the next section, we will discuss how accurately we can measure the
endpoints and the stau production cross section at the ILC.  Then, we
combine the information about the endpoints and the cross sections to
determine the stau masses.

\section{Reconstruction of Stau Properties}\label{sec:stau}

\begin{table}[t]
\centering
\renewcommand{\arraystretch}{1.5}
\rowcolors{2}{white}{gray!15}
\addtolength{\tabcolsep}{3pt}
    \caption{
    The benchmark mass spectrum for this study (the BLR1 benchmark point of Ref.~\cite{Endo:2021zal}).
    The mass parameters are in units of GeV.
    }
    \label{table:benchmark}
    \begin{tabular}{ccccc ccccc}
  \bhline{1 pt}
      $m_{\tilde{e}\w 1}$ &
      $m_{\tilde{e}\w 2}$ &
      $m_{\tilde{\mu}_1}$ &
      $m_{\tilde{\mu}_2}$ &
      $m_{\tilde{\tau}_1}$ &
      $m_{\tilde{\tau}_2}$ &
                         \makebox[2.7em]{$m_{\tilde{\chi}^0_1}$}       &
      \multicolumn{2}{c}{\makebox[2.7em]{$\cos \theta_{\tilde{\mu}}$}} &
                         \makebox[2.7em]{$\cos \theta_{\tilde{\tau}}$} \\
      155.8 & 156.7 & 154.0 & 158.5 & 113.2 & 189.8 &
                         99.3 &
      \multicolumn{2}{c}{0.631} &
                         0.703 \\
      \hline
      $m_{\rm L}$ &
      $m_{\rm R}$ &
      $M_1$ &
      $\mu$ &
      $\tan\beta$ &
      $\Omega_{\rm DM} h^2$ &
      \multicolumn{2}{c}{{$\amu[SUSY]$}} &
      \multicolumn{2}{c}{{$a_{\mu}^{(\tilde{B})}$}}
\\
      150.0& 150.0& 100.0 & 1323  & 4.94 & 0.120 &
      \multicolumn{2}{c}{$27.1\!\times\!10^{-10}$} &
      \multicolumn{2}{c}{$27.5\!\times\!10^{-10}$}
\\
\bhline{1 pt}
\end{tabular}
\end{table}

We investigate the ``BLR1 benchmark point'' of Ref.~\cite{Endo:2021zal} for the analysis.
The mass spectrum of this point is shown in Table~\ref{table:benchmark}.
This spectrum is obtained by assuming the universality of the SUSY breaking slepton masses: $m_{\tilde{L}_1}=m_{\tilde{L}_2}=m_{\tilde{L}_3}\equiv m_{\rm L}$ and $m_{\tilde{E}_1}=m_{\tilde{E}_2}=m_{\tilde{E}_3}\equiv m_{\rm R}$,
where $m_{\tilde{L}_i}$ and $m_{\tilde{R}_i}$ are soft SUSY breaking mass parameters of the left- and right-handed sleptons in the $i$-th generation, respectively.
Furthermore,
$m_{\rm L} =m_{\rm R}$ is set;\footnote{%
It is hard to probe a parameter region of $m_{\rm L} \simeq m_{\rm R}$ at the LHC when mass differences between the sleptons and the LSP are close and all the emitted leptons are soft.
}
because of this relation, the slepton mixing angles become close to $\pi/4$ (or $3\pi/4$) in the BLR1 benchmark point.
Superparticles other than the sleptons, the Bino, and the Higgsinos are assumed to be decoupled. The trilinear couplings of sleptons are set to be zero.\footnote{%
Note that the value of $\tan \beta$ is slightly different from Ref.~\cite{Endo:2021zal} by a radiative correction: $\tan \beta$ is rescaled by
$1/(1+\Delta_l)$ where $\Delta_l$ is a non-holomorphic radiative correction
 \cite{Marchetti:2008hw,Girrbach:2009uy} (see, also \cite{Kitahara:2013lfa,Endo:2013lva}).
 As a result,
  the mass eigenvalues and the mixing angles in  Table~\ref{table:benchmark} are consistent within the tree-level calculation by using the rescaled value of  $\tan\beta$.
}

At this benchmark point, the muon $g-2$ anomaly can be explained at the
1$\sigma$ level with the correct dark matter relic abundance.
Note that the value of $\amu[SUSY]$ in Table~\ref{table:benchmark} {also} includes
the Higgsino  contributions, while $\amubino$ includes only the Bino-smuon contribution.
They are calculated with the exact formula given in Ref.~\cite{vonWeitershausen:2010zr}.
It includes two-loop photonic contributions, which has a large logarithmic factor and can be sizable~\cite{Degrassi:1998es,vonWeitershausen:2010zr}.
The constraints from the LHC Run~2 data, the dark matter direct detection,
and the vacuum meta-stability condition are also satisfied~\cite{Endo:2021zal}.

We generated SUSY event samples at $\sqrt{s} = 500\GeV$ using \texttt{WHIZARD\,2.8.5}~\cite{Whizard} and stored in the mini-DST data format~\cite{kawada2021minidst}.
Tau dacays are processed with \texttt{TAUOLA\,2.7}~\cite{Chrzaszcz:2016fte}.
The detector simulation is performed with \texttt{DELPHES\,3.5.0}~\cite{DELPHES} using the ILC generic detector card~\cite{ILCDelphes}.
We included all $2f$, $4f$, $5f$, $6f$ and Higgs event samples of the SM background fully simulated in the International Large Detector (ILD)~\cite{ILDIDR} and two-photon scattering process ($\gamma \gamma \to $ two fermions)
simulated with the  \texttt{SGV} fast detector simulator~\cite{SGV}. For the full simulation samples, the PandoraPFA algorithm\cite{pandora} is used to reconstruct particles from tracks and calorimeter clusters, which are used as input for the tau reconstruction.
For tau reconstruction, we use  the \texttt{TaJetClustering} processor~\cite{TauFinder}, which clusters particles in the narrow angles consistent with tau mass as tau candidates and removes jet-like non-isolated particles.
We use two different beam polarizations:
$P_{e^-} = -80{\%}$, $P_{e^+} = +30{\%}$ and
$P_{e^-} = +80{\%}$, $P_{e^+} = -30{\%}$, which are denoted as
eLpR and
eRpL, respectively.
Based on the ILC running scenario~\cite{ILCOperatingScenario, 250GeVRun}, we assume a total integrated luminosity ${\cal L}$ of $1.6 \iab$ for each of eLpR and eRpL beam polarizations, with which the expected number of SUSY events without any cuts is $\mathcal{O}(10^5\text{--}10^6)$  while the SM background events is $\mathcal{O}(10^9)$.

\subsection{Event Selection}

In this work, we consider the following stau pair-production events,
\begin{equation}
    e^+ e^- \to \tilde{\tau}_i^\ast \tilde{\tau}_j \to \tau^+ \tau^- + 2 \tilde{B}
    ~~~
  (i,j=1,2)\,,
\end{equation}
which have two oppositely-charged  taus and large missing momentum with little other activity.
We apply the following preselections:
\begin{itemize}
    \item Require exactly two reconstructed taus with opposite charge.
    \item Remove events with one or more isolated electrons or muons in the event. We only select hadronic tau decays to mainly suppress leptonic background including $\tilde{e}\tilde{e}$ and $\tilde{\mu}\tilde{\mu}$ processes.
    \item Require two reconstructed taus to have, in total, at least one photon or at least three charged particles.
    This cut is to further reject events from $\tilde{e}\tilde{e}$ and $\tilde{\mu}\tilde{\mu}$ processes.
    \item Remove events with large non-tau activity.
    Events are removed if they contain two or more tracks, or six or more neutral particles, that are not included in the tau candidate jets.
    This efficiently removes most of semi-leptonic and hadronic SM backgrounds, while leptonic events with pileup of low energy $\gamma\gamma\to$ hadrons are still accepted.
\end{itemize}
These preselections efficiently remove non-tau background, but numerous tau backgrounds such as $\gamma\gamma\to\tau\tau$, $e^+e^- \to \tau\tau$ and $e^+e^- \to W^+W^- \to \tau \nu \tau \nu$ remain, requiring strong kinematic cuts for signal selection. Significant amount of $\tilde{e}\tilde{e}$ and $\tilde{\mu}\tilde{\mu}$ SUSY background still remains, which is due to mis-identification of electrons and muons.
The lepton identification criteria can be further optimized after reproducing SUSY events with the full detector simulation.

We apply the following kinematic cuts to further reduce backgrounds:
\begin{itemize}
    \item Cut 1: $\theta _{\mathrm{acop}} / \pi > 0.05$
    \item Cut 2: $20 < E_{\mathrm{vis}} < 300$~GeV
    \item Cut 3: $M_{\mathrm{inv}} > 200$~GeV
    \item Cut 4: $|\cos \theta _{\mathrm{miss}}| < 0.9$
    \item Cut 5: missing $P_t > 20$~GeV
    \item Cut 6: $|\cos \theta _{\tau ^{\pm}}| < 0.9$
\end{itemize}
where $\theta _{\mathrm{acop}}$ is the acoplanarity angle between two reconstructed taus, $E_{\mathrm{vis}}$ is the visible energy of an event, $M_{\mathrm{inv}}$ is the missing mass of an event, $\theta _{\mathrm{miss}}$ is the polar angle of missing momentum, and $\theta _{\tau ^{\pm}}$ is the polar angle of $\tau ^{\pm}$.
{Tables~\ref{tab:cuttable_eLpR} and~\ref{tab:cuttable_eRpL} show the cut tables for the eLpR and eRpL beam polarizations, respectively.
For both beam configurations, $\mathcal{O}(10^3 \text{--} 10^4)$ stau signal and $\mathcal{O}(10^4)$
background events remain after all the selection cuts.
As one can see from the tables,
the SM background
in the eLpR beam polarization is enhanced by
the weak-boson exchanging processes (\eg, $e^+e^-\to W^+W^- \to \tau\nu\tau\nu$).
On the other hand,
the SUSY background
is amplified in the eRpL one by the
$t$-channel Bino-exchange process $e^+e^- \to \tilde{e}^{\ast}_{\rm R} \tilde{e}_{\rm R}$.}

\begin{table}[p]
    \centering
    \renewcommand{\arraystretch}{1.5}
    \newcommand\TIMES{\!\times\!}
    \rowcolors{2}{gray!15}{white}
    \caption{\label{tab:cuttable_eLpR}%
    The cut table {for the eLpR beam polarization}.
    An integrated luminosity ${\cal L}=1.6 \iab$ is assumed.
    $\tilde{\tau}_1\tilde{\tau}_2$ stands for $\tilde{\tau}_1^\ast \tilde{\tau}_2 +\tilde{\tau}_2^\ast \tilde{\tau}_1 $ events.
    The $\gamma\gamma\to2f$ events are not included in the ``SM bkg'' but in a separate column.
    }
    \scalebox{0.96}{
    \begin{tabular}{ccccccc}
    \bhline{1pt}
     & $\tilde{\tau}_1 \tilde{\tau}_1$ & $\tilde{\tau}_2 \tilde{\tau}_2$ & $\tilde{\tau}_1 \tilde{\tau}_2$ & SUSY bkg & SM bkg & $\gamma \gamma \to 2f$ \\ \hline
    no cuts & $1.488 \TIMES 10^5$ & $4.647 \TIMES 10^4$ & $2.621 \TIMES 10^4$ & $5.539 \TIMES 10^5$ & $8.770 \TIMES 10^7$ & $4.283 \TIMES 10^9$ \\
    preselection & $2.157 \TIMES 10^4$ & $1.340 \TIMES 10^4$ & 5176 & 4653 & $1.209 \TIMES 10^5$ & $3.047 \TIMES 10^7$ \\
    cut 1 & $1.703 \TIMES 10^4$ & $1.230 \TIMES 10^4$ & 4536 & 4284 & $4.131 \TIMES 10^4$ & $1.310 \TIMES 10^7$ \\
    cut 2 & $1.608 \TIMES 10^4$ & $1.229 \TIMES 10^4$ & 4499 & 4284 & $2.585 \TIMES 10^4$ & $7.058 \TIMES 10^6$ \\
    cut 3 & $1.608 \TIMES 10^4$ & $1.229 \TIMES 10^4$ & 4499 & 4284 & $2.080 \TIMES 10^4$ & $6.069 \TIMES 10^6$ \\
    cut 4 & $1.475 \TIMES 10^4$ & $1.141 \TIMES 10^4$ & 4141 & 3882 & $1.368 \TIMES 10^4$ & $5.963 \TIMES 10^5$ \\
    cut 5 & 4798 & $1.091 \TIMES 10^4$ & 3675 & 3760 & $1.151 \TIMES 10^4$ & 2788 \\
    cut 6 & 4456 & 9457 & 3397 & 2961 & 7681 & 1764 \\
    \bhline{1pt}
   \end{tabular}
   }

  \vspace{1cm}

    \rowcolors{2}{gray!15}{white}
    \caption{\label{tab:cuttable_eRpL}%
    The cut table for the eRpL beam polarization.
    An integrated luminosity ${\cal L}=1.6 \iab$ is assumed.
    }
    \scalebox{0.96}{
    \begin{tabular}{ccccccc}
    \bhline{1pt}
     & $\tilde{\tau}_1 \tilde{\tau}_1$ & $\tilde{\tau}_2 \tilde{\tau}_2$ & $\tilde{\tau}_1 \tilde{\tau}_2$ & SUSY bkg & SM bkg & $\gamma \gamma \to 2f$ \\ \hline
    no cuts & $1.386 \TIMES 10^5$ & $4.211 \TIMES 10^4$ & $2.075 \TIMES 10^4$ & $1.286 \TIMES 10^6$ & $4.727 \TIMES 10^7$ & $4.283 \TIMES 10^9$ \\
    preselection & $2.004 \TIMES 10^4$ & $1.213 \TIMES 10^4$ & 4128 & $1.380 \TIMES 10^4$ & $7.292 \TIMES 10^4$ & $3.047 \TIMES 10^7$ \\
    cut 1 & $1.581 \TIMES 10^4$ & $1.113 \TIMES 10^4$ & 3616 & $1.268 \TIMES 10^4$ & $1.916 \TIMES 10^4$ & $1.310 \TIMES 10^7$ \\
    cut 2 & $1.493 \TIMES 10^4$ & $1.112 \TIMES 10^4$ & 3584 & $1.268 \TIMES 10^4$ & 8032 & $7.058 \TIMES 10^6$ \\
    cut 3 & $1.493 \TIMES 10^4$ & $1.112 \TIMES 10^4$ & 3584 & $1.268 \TIMES 10^4$ & 4954 & $6.069 \TIMES 10^6$ \\
    cut 4 & $1.369 \TIMES 10^4$ & $1.032 \TIMES 10^4$ & 3301 & $1.154 \TIMES 10^4$ & 2119 & $5.963 \TIMES 10^5$ \\
    cut 5 & 4396 & 9868 & 2930 & $1.117 \TIMES 10^4$ & 1439 & 2788 \\
    cut 6 & 4091 & 8564 & 2706 & 8940 & 1001 & 1764 \\
    \bhline{1pt}
   \end{tabular}
   }
    \addtolength{\tabcolsep}{-3pt}
\end{table}

We further assume that the background events from $\gamma \gamma \to 2f$ processes can be greatly reduced by BeamCal information~\cite{Bechtle:2009em}.
As the number of those events is already subdominant
after the six cuts,
we will simply omit the $\gamma\gamma\to2f$ background in the further analysis.

\subsection{Measurements of the Cross Sections and Endpoints}

After the selection cuts, each event has two reconstructed taus.
Our analysis uses the reconstructed energy
of the more energetic
$\tau$, which we denote by $E_\tau$.
The distribution of $E_\tau$
  is shown in Figs.~\ref{fig:hightau_E_eLpR_N} and
  \ref{fig:hightau_E_eRpL_N} for the eLpR and eRpL beam
  configurations, respectively.  From the plots, we extract endpoints
  and number of events, which are necessary to reconstruct stau masses
  $m_{\tilde{\tau}_1}$, $m_{\tilde{\tau}_2}$ and the mixing angle
  $\theta _{\tilde{\tau}}$.
We assume that the masses of
$\tilde{e}$ and $\tilde{\mu}$ are obtained by similar but
independent analyses.  Since the endpoints in association with
  two staus are well separated from those associated to
  $\tilde e$ and $\tilde \mu$ in the present case, they are not
affected by the SUSY background so much.  Note that, throughout our analysis,
we include only the statistical uncertainty. No systematic
uncertainty, including one coming from the MC statistics, is assigned
for simplicity.

\subsubsection{Measurements with eLpR Polarization}

\begin{figure}[p]
  \centering
  \includegraphics[width=0.8\linewidth]{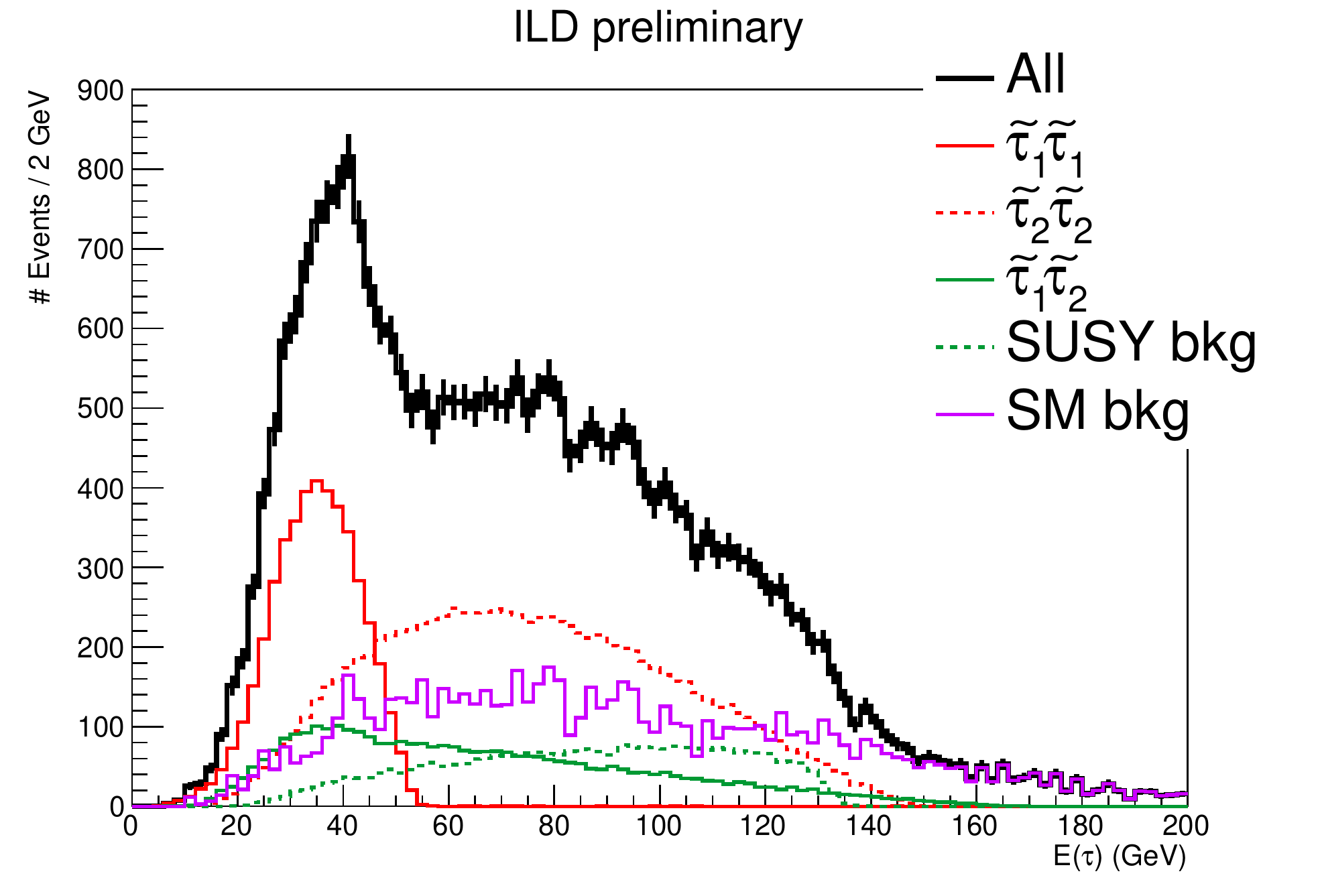}
  \caption{ The reconstructed higher $\tau$ energy distribution in an event after applying all selection cuts for eLpR beam polarization.
  The vertical error bars correspond to the expected statistical uncertainty in the actual running.}
  \label{fig:hightau_E_eLpR_N}
\vspace{1cm}
  \centering
  \begin{minipage}[b]{0.49\linewidth}
    \centering
    \includegraphics[width=1.0\linewidth,trim=8 0 8 0]{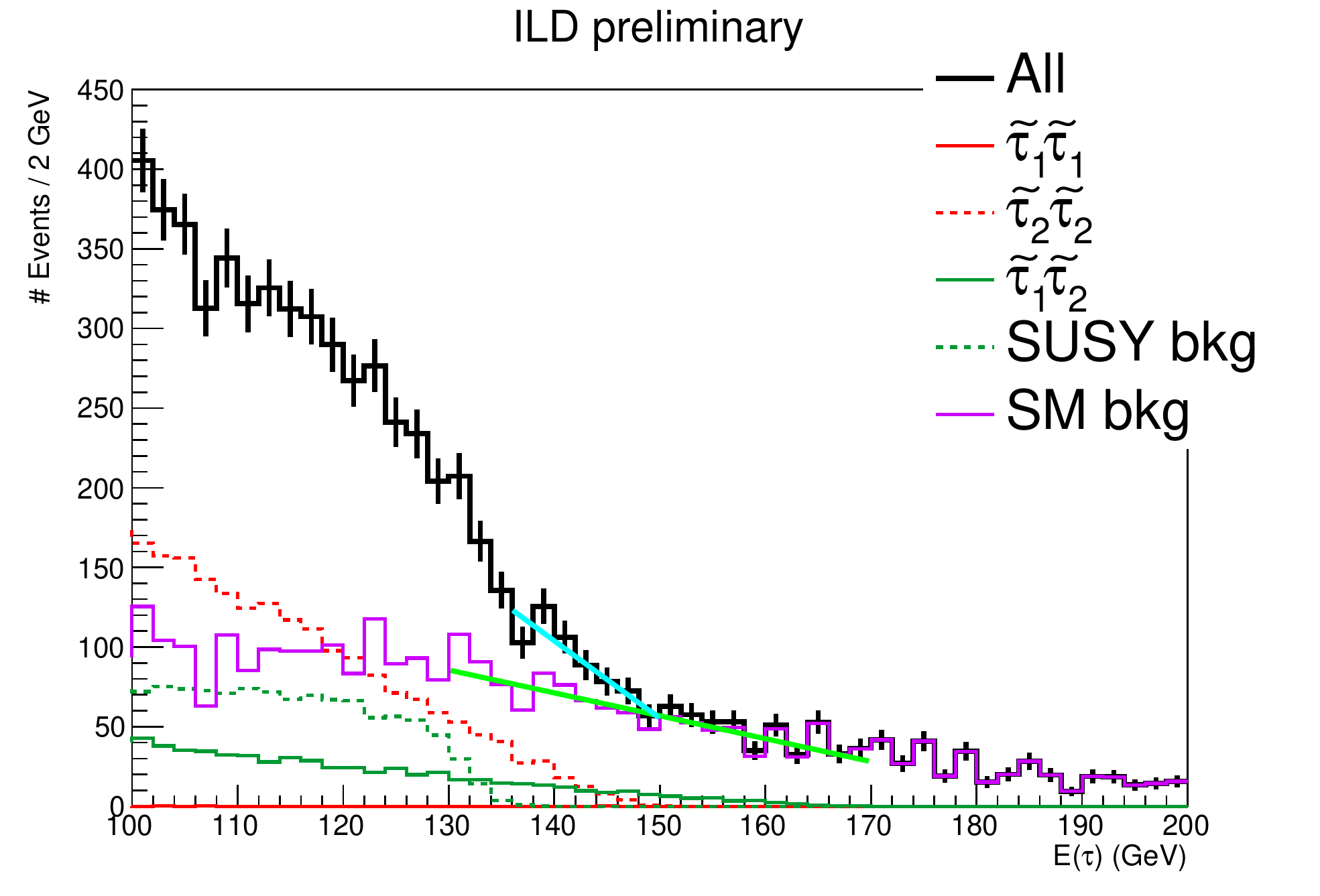}
    (a) $\tilde{\tau}_2\tilde{\tau}_2$ endpoint fit
  \end{minipage}
  \begin{minipage}[b]{0.49\linewidth}
    \centering
    \includegraphics[width=1.0\linewidth,trim=8 0 8 0]{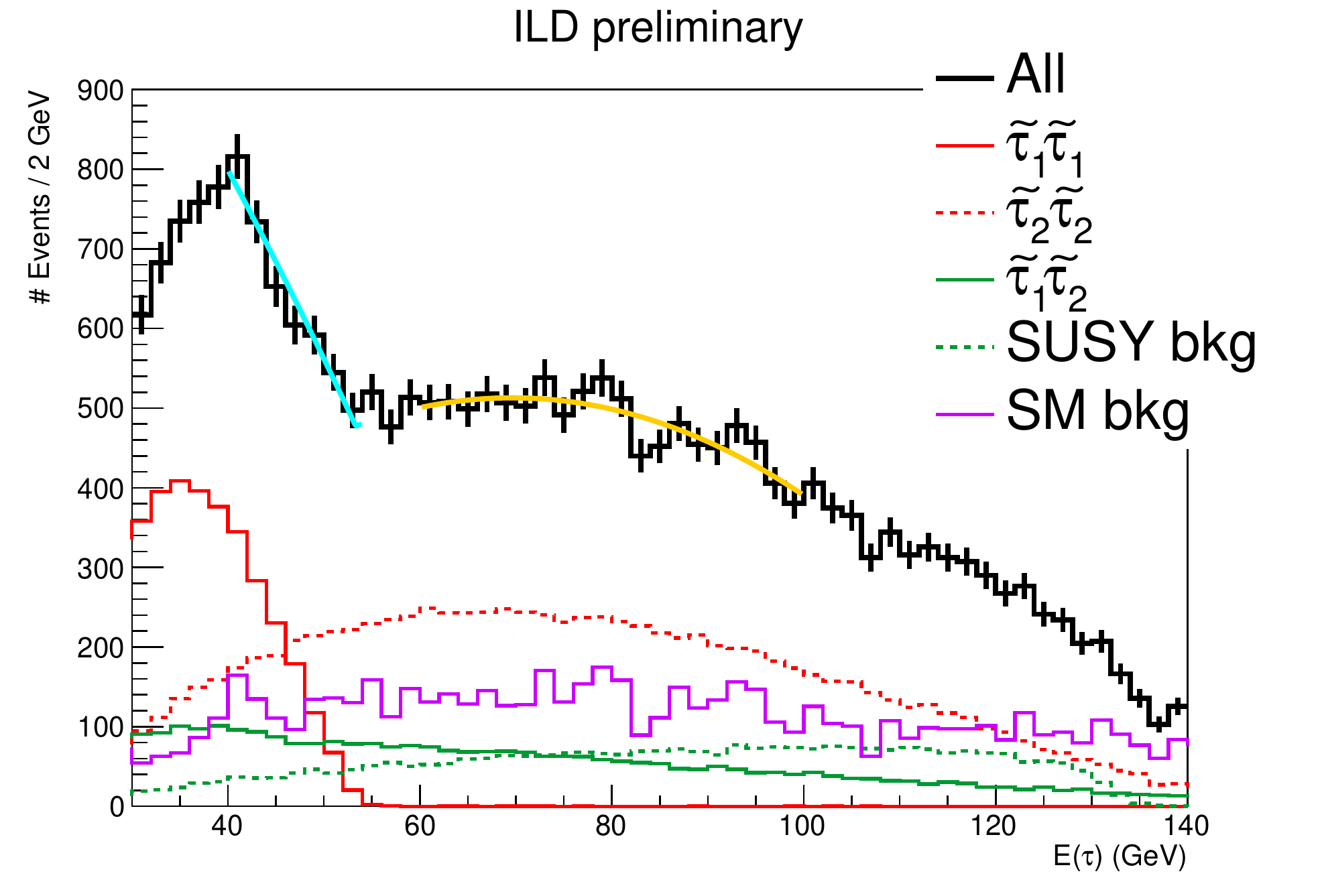}
    (b) $\tilde{\tau}_1\tilde{\tau}_1$ endpoint fit
  \end{minipage}
  \caption{
  Zoomed regions of Fig.~\ref{fig:hightau_E_eLpR_N}.
  (a) The $\tilde{\tau}_2\tilde{\tau}_2$ endpoint region.
  The light green and cyan lines correspond to the
  SM background fit
  and the fit of all the events, respectively.
  (b) The $\tilde{\tau}_1\tilde{\tau}_1$ endpoint region.
  The cyan (yellow) line is the fit result {of all the events}
  for the $E_\tau$ range of 40--$54\GeV$ (60--$100\GeV$).
  See text for details of the fittings.
  }
  \label{fig:stau1122fit_eLpR}
\end{figure}

Let us begin with the eLpR beam configuration.
We first
extract the endpoint $E_{+} (\tilde{\tau}_2)$ from
the pair-production process
$e^+e^-\rightarrow\tilde{\tau}_2^*\tilde{\tau}_2$.  In
Fig.~\ref{fig:stau1122fit_eLpR} (a), we show the energy distribution
of $E_\tau$ as well as the results of the fits in the signal and
background regions.  In the present case, the effect of the SUSY backgrounds is expected to be small because, at $E_\tau\sim E_{+} (\tilde{\tau}_2)$, the SUSY event is dominated by the $\tilde{\tau}_2^*\tilde{\tau}_2$ pair-production event.  On the contrary, the SM background is substantial. We assume that, at the time of the ILC operation, the SM background for the extraction of $E_{+} (\tilde{\tau}_2)$ will be well understood by combining the side-band data and the MC analysis and that the endpoint can be determined by comparing the real data with the SM background.  In the present analysis, the would-be-known SM background is estimated from the MC result shown in Fig.~\ref{fig:stau1122fit_eLpR} (a); we use the SM background distribution given 
by the straight line determined from our MC data in the range of 130--170~GeV (the light green
line).  We then
perform fitting for all the events using an additional straight line
on the SM fit in the range of 136--150~GeV (the cyan line).  We can
extract the endpoint as the crossover of these lines.  Then, we
obtain
\begin{equation}
    E_{+}(\tilde{\tau}_2) = 149.5 \pm 1.7\GeV\,,
        \label{eq:end2_eLpR}
\end{equation}
consistent with the theoretical endpoint value 149.9~GeV.

The $\tilde{\tau}_1$ endpoint is also extracted as shown in Fig.~\ref{fig:stau1122fit_eLpR} (b).
We first fit the distribution of all the events using a second-order polynomial in the range of 60--100~GeV (the yellow line) for the estimation of the background.
Here we do not assume that the contributions from the $\tilde{\tau}_1\tilde{\tau}_2$ and $\tilde{\tau}_2\tilde{\tau}_2$ processes are known, since they depend on the yet-unknown stau mixing parameter.
This treatment allows to regard this procedure as a data-driven estimation of the background.
We then in the range 40--54~GeV perform a fit (the cyan line) by a linear function
stacked
on the extrapolation of the aforementioned second-order polynomial fit function.
The obtained endpoint for $\tilde{\tau}_1$ is
\begin{equation}
    E_{+}(\tilde{\tau}_1) = 53.31 \pm 0.55\GeV\,,
    \label{eq:end1_eLpR}
\end{equation}
which is smaller than the theoretical endpoint 54.5~GeV.

For extracting the stau cross section,
we use the number of events in the energy range $60\GeV < E_{\tau} < 150\GeV$.
For an integrated luminosity ${\cal L}=1.6\iab$ for eLpR beam polarization, the expected number
of signal and background events in this range are found to be
\begin{align}
    N_{22} &= 6413\,, \label{eq:N22_eLpR} \\
    N_{12} &= 1705\,, \\
    N_{\text{SM}} &= 4873\,,    \label{eq:NSMbkg_eLpR} \\
    N_{\text{SUSY}} &= 2365\,,
    \label{eq:NSUSYbkg_eLpR}
\end{align}
where $N_{22}$, $N_{12}$, $N_{\text{SM}}$, and $N_{\text{SUSY}}$ are the numbers of events of $\tilde{\tau}_2\tilde{\tau}_2$ production, $\tilde{\tau}_1\tilde{\tau}_2$ production, SM background, and SUSY background, respectively.

\subsubsection{Measurements with eRpL Polarization}

\begin{figure}[p]
  \centering
  \includegraphics[width=0.8\linewidth]{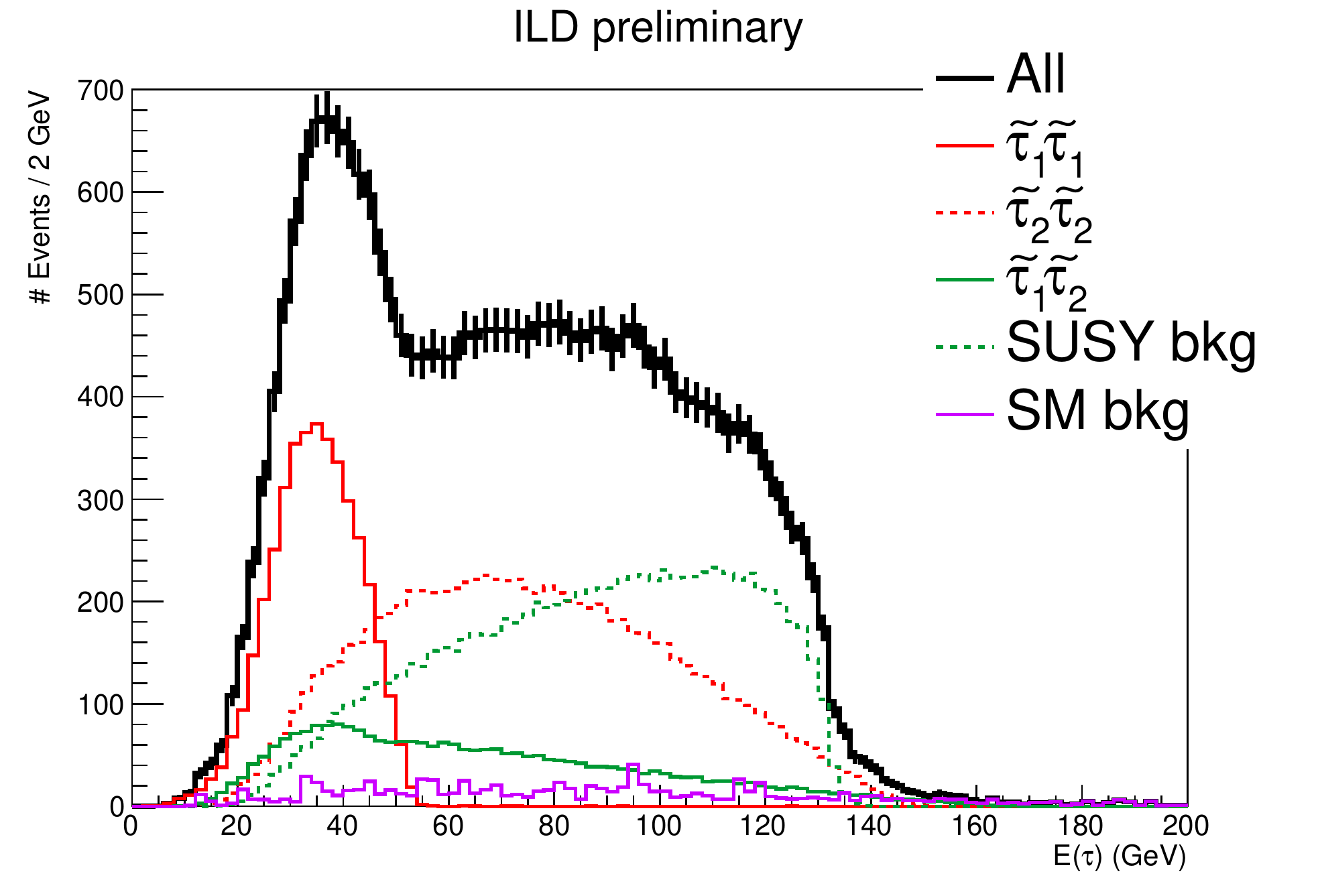}
  \caption{The reconstructed higher $\tau$ energy distribution in an event after applying all selection cuts for eRpL beam polarization.
  The vertical error bars correspond to the expected statistical uncertainty in the actual running.}
  \label{fig:hightau_E_eRpL_N}

  \vspace{1cm}

  \begin{minipage}[b]{0.49\linewidth}
    \centering
    \includegraphics[width=1.0\linewidth,trim=8 0 8 0]{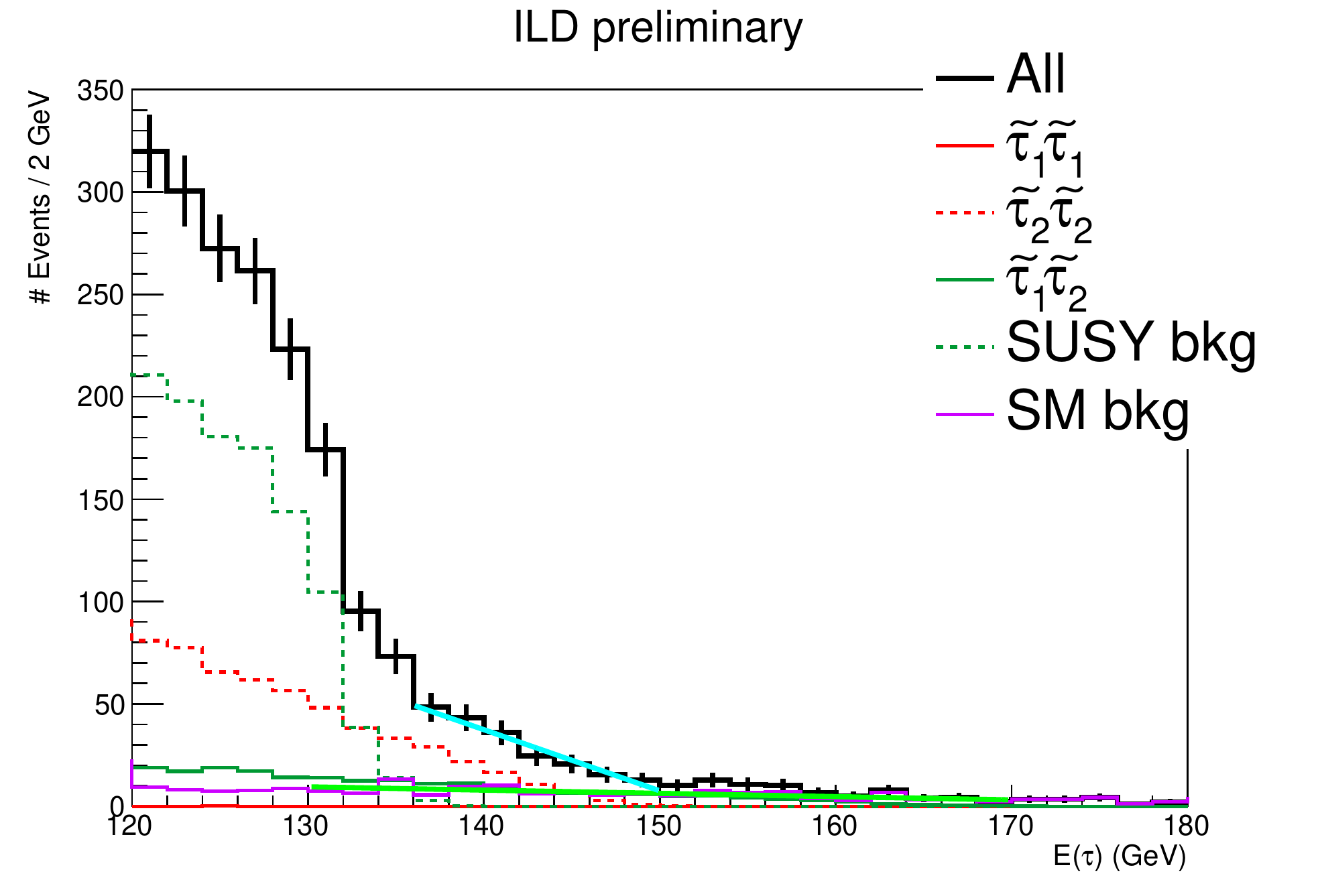}
    (a) $\tilde{\tau}_2\tilde{\tau}_2$ endpoint fit
  \end{minipage}
  \begin{minipage}[b]{0.49\linewidth}
    \centering
    \includegraphics[width=1.0\linewidth,trim=8 0 8 0]{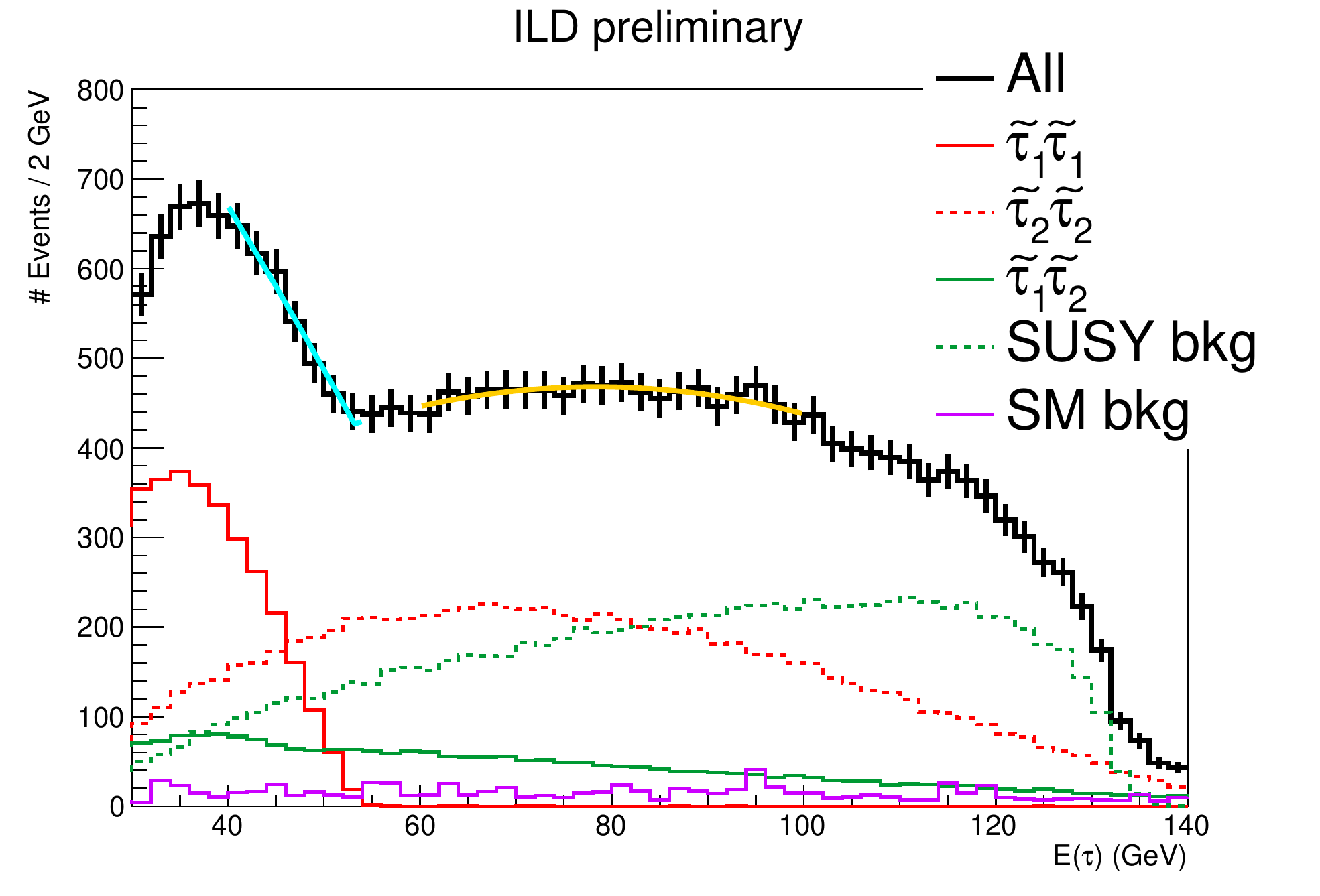}
    (b) $\tilde{\tau}_1\tilde{\tau}_1$ endpoint fit
  \end{minipage}
  \caption{Zoomed regions of Fig.~\ref{fig:hightau_E_eRpL_N}.
  (a) The $\tilde{\tau}_2\tilde{\tau}_2$ endpoint region.
  The light green and cyan lines correspond to the
  SM background fit
  and the fit of all the events, respectively.
  (b) The $\tilde{\tau}_1\tilde{\tau}_1$ endpoint region.
  The cyan (yellow) line is the fit result {of all the events}
  for the $E_\tau$ range of 40--$54\GeV$ (60--$100\GeV$).
  See text for details of the fittings.
  }
  \label{fig:stau1122fit_eRpL}
\end{figure}

We repeat the aforementioned analysis for the eRpL beam configuration.
The resulting $E_\tau$ distribution is shown in Fig.~\ref{fig:hightau_E_eRpL_N}.
As displayed in Fig.~\ref{fig:stau1122fit_eRpL}, the $\tilde\tau_2$ endpoint is obtained as
\begin{equation}
    E_{+}(\tilde{\tau}_2) = 150.4 \pm 1.2\GeV\,,
        \label{eq:end2_eRpL}
\end{equation}
which is consistent with the theoretical endpoint value 149.9~GeV, and the $\tilde\tau_1$ endpoint is
\begin{equation}
    E_{+}(\tilde{\tau}_1) = 53.17 \pm 0.67\GeV\,,
    \label{eq:end1_eRpL}
\end{equation}
which is slightly smaller than the theoretical endpoint 54.5~GeV.

For an integrated luminosity ${\cal L}=1.6\iab$ for eRpL beam polarization, the expected number of signal and background events in the range {$60\GeV < E_{\tau} < 150\GeV$} are found to be
\begin{align}
    N_{22} &= 5803\,, \label{eq:N22_eRpL} \\
    N_{12} &= 1354\,, \\
    N_{\text{SM}} &= 595.2\,,    \label{eq:NSMbkg_eRpL} \\
    N_{\text{SUSY}} &= 7215\,.
    \label{eq:NSUSYbkg_eRpL}
\end{align}

\subsection{Fit of Stau Masses and Mixing Angle}

{In this section,
we fit the stau masses and the mixing angle from the values of the endpoints and the expected number of signal events simultaneously.
We define the following $\chi^2$ variable:
}
\begin{equation}
    \chi^2_{\mathrm{tot}} (m_{\tilde{\tau}_1},m_{\tilde{\tau}_2}, \cos \theta_{\tilde{\tau}}, m_{\tilde{\chi}_1^0}) =
    \chi^2_{m_{\tilde{\chi}_1^0}}+\sum_{\text{pol.}}\left[ \chi^2_{E_+(\tilde{\tau}_1)}+\chi^2_{E_+(\tilde{\tau}_2)} + \chi^2_{N(\tilde{\tau})}\right]
\,,
\end{equation}
{where the polarization (pol.) sum runs over the eLpR and eRpL polarization configurations and
$\chi^2_{\mathcal{O}}$ for each observable $\mathcal{O}$ is given by
\begin{align}
  \chi^2_{\mathcal{O}} \equiv
  \frac{(\mathcal{O}^{\rm th} - \mathcal{O}^{\rm exp})^2}
       {\sigma_{\mathcal{O}}^2}
\end{align}
with $\mathcal{O}^{\rm th}$ and $\mathcal{O}^{\rm exp}$ being theoretically predicted and experimentally observed values, respectively, and $\sigma_{\mathcal{O}}^2$ being the variance.
Here, $\chi^2_{m_{\tilde{\chi}_1^0}}$ and
$\chi^2_{N(\tilde{\tau})}$ are for the LSP mass and
the number of the stau events, respectively.
We neglect
the correlation among different observables.}  Using the $\chi^2$
variable, the probability distribution function (PDF) is defined as ${\cal
  P}\propto e^{-\chi^2_{\mathrm{tot}}/2}$.

For the calculation of $\chi^2_{m_{\tilde{\chi}_1^0}}$, we use
$m_{\tilde{\chi}^0_1}^{\rm exp} = 99.3\GeV$
and $\sigma_{m_{\tilde{\chi}^0_1}} = 0.1\GeV$,
where the former comes from the true value in the model point (see Table~\ref{table:benchmark}) and the latter, associated to an endpoint search in the
process $e^+ e^- \to \tilde{\mu}^+\tilde{\mu}^-/\tilde{e}^+\tilde{e}^-$, is adopted from Refs.~\cite{Martyn:2004jc,Endo:2013xka}.
Theoretical values
of the endpoints
for $\chi^2_{E_+(\tilde{\tau}_1)}$ and $\chi^2_{E_+(\tilde{\tau}_2)}$
are given by Eq.~\eqref{eq:endpoint} with $E_{\tilde{\tau}_i} =\sqrt{s}/2$, while
$E_{+}(\tilde{\tau}_i)^{\rm exp}$'s are taken from Eqs.~\eqref{eq:end1_eLpR} and \eqref{eq:end1_eRpL} for $i=1$ and Eqs.~\eqref{eq:end2_eLpR} and \eqref{eq:end2_eRpL} for $i=2$.

For the calculation of
$\chi^2_{N(\tilde{\tau})}$, we utilize the number of events, Eqs.~\eqref{eq:N22_eLpR}--\eqref{eq:NSUSYbkg_eLpR} and Eqs.~\eqref{eq:N22_eRpL}--\eqref{eq:NSUSYbkg_eRpL} for eLpR and eRpL polarization configurations, respectively.
There, we imposed an additional cut, $60\GeV <  E_{\tau} < 150\GeV$, to collect $\tilde{\tau}_1\tilde{\tau}_2$ and $\tilde{\tau}_2\tilde{\tau}_2$ events.
This cut eliminates $\tilde{\tau}_1\tilde{\tau}_1$ events and makes the number of events more sensitive to the stau mixing angle.
We construct $\chi^2_{N(\tilde{\tau})}$ with
\begin{align}
N(\tilde{\tau})^{\rm th} &= \mathcal{L} \left[2 \epsilon_{12} \,  \sigma(e^+ e^- \to \tilde{\tau}_1^\ast \tilde{\tau}_2) +  \epsilon_{22} \, \sigma(e^+ e^- \to \tilde{\tau}_2^\ast \tilde{\tau}_2) \right]\,,
\\
N(\tilde{\tau})^{\rm exp} & = N_{\rm total} - N_{\rm bkg}\,,
\\
\sigma_{N(\tilde{\tau})} &= \sqrt{N_{\rm total} + N_{\rm bkg}}\,.
\end{align}
Here, we first evaluate the signal efficiencies
($\epsilon_{12}$ and $\epsilon_{22}$) by comparing the MC result in the range of $60\GeV < E_{\tau} < 150\GeV$ with the theoretical cross section in Eq.~\eqref{eq:CrossSection}.
It is assumed that the signal efficiencies are insensitive to the stau mixing angle and masses.
The number of total events in the range of $60\GeV < E_{\tau} < 150\GeV$ is $N_{\rm total}$, which will be given by experiment, and we use
$N_{\rm total} = N_{\rm signal} + N_{\rm bkg}$ and $ N_{\rm signal} = N_{12} + N_{22}$.
The number of background events, $N_{\rm bkg}=N_{\text{SUSY}} + N_\text{SM}$,
will be determined by the MC analysis.

Note that $\chi^2_{N(\tilde{\tau})}$  has
a twofold ambiguity with regard to the stau mixing angle $\theta_{\tilde{\tau}}$.
Since the slepton pair-production cross section in Eq.~\eqref{eq:CrossSection} is invariant under
$\theta_{\tilde{l}} \leftrightarrow (\pi-\theta_{\tilde{l}})$, we cannot  resolve this degeneracy.
This degeneracy corresponds to the sign of the $\mu$ parameter.
The sign of the muon $g-2$ anomaly implies $\mu M_1 >0 $. Therefore, this degeneracy also corresponds to the sign of $M_1$.
In our fitting, we assumed $\mu>0$ with $M_1>0$.

Using the PDF, ${\cal
  P}\propto e^{-\chi^2_{\mathrm{tot}}/2}$, we obtain
  \begin{align}
m_{\tilde{\tau}_1} &=  112.8 \pm 0.2 \GeV\,,
\label{eq:stau1fit}\\
m_{\tilde{\tau}_2} &=
189.9 _{-0.7}^{+0.8}\GeV\,,
\label{eq:stau2fit}\\
\cos  \theta_{\tilde{\tau}}
&= 0.703 \pm0.010\,,
\end{align}
with $68\%$ probability ranges.
Here, we require that both edges of the uncertainty have the same value of the PDF.
From Eq.~\eqref{eq:sin2thl}, the stau left-right mixing parameter is given by
\begin{equation}
     - m_{\tilde\tau{LR}}^2 =
      \left(1.17 \pm 0.01 \right)\times 10^{4}\GeV{}^2\,.
      \label{eq:LRstau}
\end{equation}
Then the smuon left-right mixing parameter can be reconstructed from Eq.~\eqref{eq:mlrratio} as
\begin{equation}
   - m_{\tilde\mu{LR}}^2 = 693_{-8}^{+9} \GeV{}^2\,.
    \label{eq:LRsmu}
\end{equation}
All the fitted values {and the corresponding theoretical values at the benchmark point} are summarized in Table~\ref{table:fitresult}.

\begin{table}[t]
\centering
\renewcommand{\arraystretch}{1.5}
\rowcolors{2}{gray!15}{white}
\addtolength{\tabcolsep}{3pt}
    \caption{\label{table:fitresult}%
    Fitting summary for the stau parameters with $\sqrt{s}=500\GeV$ and ${\cal L}=1.6\iab$.
    The mass parameters are in units of GeV
and the left-right mixing parameter is in units of GeV${}^2$.}
    \begin{tabular}{ccccc}
      \bhline{1 pt}&
      $m_{\tilde{\tau}_1}$ &
      $m_{\tilde{\tau}_2}$ &
      $\cos \theta_{\tilde{\tau}}$ &
      $-m_{\tilde{\tau}{LR}}^2$ \\
      \hline
       Theoretical value &
       113.2 & 189.8 & 0.703   & 11606 \\
       Fit result&
       $112.8 \pm 0.2$&
       $189.9 _{-0.7}^{+0.8}$ &
       $0.703 \pm 0.010$&
      $\left(1.17 \pm 0.01 \right)\!\times\! 10^{4}$
\\
\bhline{1 pt}
\end{tabular}
\addtolength{\tabcolsep}{-3pt}
\end{table}

\section{Reconstruction of the Muon \texorpdfstring{\boldmath{$g-2$}}{g-2}}
\label{sec:amu}

In this section, we study an implication of the determinations of
stau parameters, in particular $m_{\tilde{\tau}{LR}}^2$, for the muon
$g-2$ anomaly.  We discuss how accurately the SUSY contribution to
$\amu$ is determined using the ILC measurements at our benchmark point.  It is assumed
that all the charged sleptons and the lightest neutralino (as the LSP)
are within the kinematical reach of the ILC
and that no other superparticles, in particular
charginos, are discovered.  In such a case, the neutralino can be
considered as the Bino-dominated one.  Furthermore, in the absence of the chargino discovery, we may expect that the SUSY
contribution to the muon $g-2$ is dominated by the Bino-smuon loop contribution $a_\mu^{(\tilde{B})}$.

We can reconstruct $a_\mu^{(\tilde{B})}$ once the following quantities are known:\footnote{For simplicity, we do not consider the determination of the lepton-slepton-Bino couplings, and assume that the coupling satisfies the tree-level SUSY relation. See Ref.~\cite{Endo:2013xka}.}
\begin{itemize}
\item Smuon mass eigenvalues: $m_{\tilde{\mu}_1}$ and $m_{\tilde{\mu}_2}$,
\item Smuon left-right mixing parameter: $m_{\tilde{\mu}{LR}}^2$,
\item The lightest neutralino mass: $m_{\tilde{\chi}^0_1}$.
\end{itemize}
At the ILC, the smuon masses can be measured from the smuon pair-production processes, and the lightest neutralino mass can be done from the selectron production processes in addition to the smuon ones.
As for the smuon left-right mixing parameter $m_{\tilde{\mu}{LR}}^2$, as we have discussed,
the stau study at the ILC can be used for its determination.  It is obtained from the stau left-right mixing parameter (see Eq.~\eqref{eq:mlrratio}).\footnote{%
  At the one-loop level, non-holomorphic corrections to the lepton
  masses may be sizable.  The non-holomorphic corrections are
  parameterized by $\Delta_l$-parameters  and Eq.\ \eqref{eq:mlrratio} is modified as \cite{Marchetti:2008hw,Girrbach:2009uy}
  \begin{align*}
    m_{\tilde{\mu}{LR}}^2
    =
    \frac{m_\mu}{m_\tau}
    \frac{1+\Delta_\tau}{1+\Delta_\mu}
    m_{\tilde{\tau}{LR}}^2\,.
  \end{align*}
  In general, $\Delta_\mu$ and $\Delta_\tau$ differ and we should
  better take account of their effects.  Importantly, $\Delta_\mu$ and
  $\Delta_\tau$ both depends only on the combination $\mu\tan\beta$
  (as well as on the masses of sleptons and Binos and Bino-muon-smuon
  couplings).  Thus, even at the one-loop level, we can obtain
  $m_{\tilde{\mu}{LR}}^2$ from $m_{\tilde{\tau}{LR}}^2$ through the
  reconstruction of $\mu\tan\beta$ in the situation of our interest.
  In our analysis,
  we neglect a difference between  $\Delta_\mu$ and $\Delta_\tau$.
  (Notice that, at the benchmark point of our choice,
  $\Delta_\mu\simeq\Delta_\tau$ holds with a good accuracy.)
}
Once $m_{\tilde{\mu}{LR}}^2$ and the smuon masses are known, the smuon mixing angle $\theta_{\tilde\mu}$ is determined from the relation,
\begin{equation}
  \sin 2 \theta_{\tilde\mu} =
  \frac{2 m_{\tilde\mu{LR}}^2}{m_{\tilde\mu_1}^2 - m_{\tilde\mu_2}^2}
  =
  \frac{m_\mu}{m_\tau}
  \frac{m_{\tilde\tau_2}^2 - m_{\tilde\tau_1}^2}{m_{\tilde\mu_2}^2 - m_{\tilde\mu_1}^2}
  \sin 2 \theta_{\tilde\tau}  \,,
  \label{SmuMixingAngle}
\end{equation}
with which the smuon mixing matrix $U_{\tilde{\mu}}$ is obtained.

With the above mentioned parameters, the Bino-smuon contribution to $\amu[SUSY]$ can be calculated.
At the
one-loop level, it is given by\footnote{This formula is exact when the lightest neutralino is composed of the Bino and the other contributions, \ie, those from the Wino and Higgsinos, are decoupled.}
\begin{align}
  a_\mu^{(\tilde{B},{\rm \ 1\mhyphen loop})} =&
  \frac{1}{16\pi^2} \sum_{A=1,2} \frac{m_\mu^2}{m_{\tilde{\mu}_A}^2}
  \left\{ -\frac{1}{12}
    \left[(\hat{N}^{\mu_L}_{A})^2 + (\hat{N}^{\mu_R}_{A})^2\right] F^N_1\left(\frac{m_{\tilde\chi^0_1}^2}{m_{\tilde{\mu}_A}^2}\right) \right.
    \nonumber \\ & \qquad\qquad\qquad\quad~~
\left. - \frac{m_{\tilde\chi^0_1}}{3 m_\mu}
    \hat{N}^{\mu_L}_{A} \hat{N}^{\mu_R}_{A} F^N_2\left(\frac{m_{\tilde\chi^0_1}^2}{m_{\tilde{\mu}_A}^2}\right)
  \right\}\,,
  \label{eq:amu_ILC}
\end{align}
where
\begin{align}
  \hat{N}^{\mu_{L}}_{A} =
  \frac{1}{\sqrt{2}} g_Y
  (U_{\tilde{\mu}})_{A1}\,,\qquad
  \hat{N}^{\mu_{R}}_{A} =
  -\sqrt{2} g_Y
  (U_{\tilde{\mu}})_{A2}\, ,
  \label{eq:NAhat}
\end{align}
with $g_Y$ being the U(1)$_Y$ gauge coupling constant,
and the loop functions are defined as
\begin{align}
  F^N_1(x) &= \frac{2}{(1-x)^4} \left( 1-6x+3x^2+2x^3-6x^2\ln x \right)\,,
  \\
  F^N_2(x) &= \frac{3}{(1-x)^3} \left( 1-x^2+2x\ln x \right)\,.
\end{align}
In our analysis, we include the photonic correction to evaluate $a_\mu^{(\tilde{B})}$ as
\begin{align}
    a_\mu^{(\tilde{B})} =
    a_\mu^{(\tilde{B},{\rm \ 1\mhyphen loop})} +
    a_\mu^{(\tilde{B},{\rm \ photonic})},
    \label{eq:amuB}
\end{align}
where $a_\mu^{(\tilde{B},{\rm \ photonic})}$ is the two-loop photonic contribution~\cite{vonWeitershausen:2010zr}.

The stau study of the ILC can give crucial information for the reconstruction
of
$a_\mu^{(\tilde{B})}$
through the determination of $m_{\tilde{\mu}{LR}}^2$.  To understand the impact of the stau study, let us estimate the uncertainty related to $m_{\tilde{\mu}{LR}}^2$ in the reconstructed value of $a_\mu^{(\tilde{B})}$.
Including only the uncertainty in Eq.~\eqref{eq:LRsmu}, we obtain
\begin{align}
 \left. a_\mu^{(\tilde{B})} \right|_{m_{\tilde{\mu}{LR}}^2} =
  (27.6 \pm 0.3) \times10^{-10}
  \quad[\pm 1\%]\,.
  \label{eq:amuB_m2LRmu}
\end{align}
With including all the uncertainties, $a_\mu^{(\tilde{B})}$ is reconstructed as
\begin{align}
 a_\mu^{(\tilde{B})} =
  (27.5 \pm 0.4) \times10^{-10}
  \quad[\pm 1\%]\,,
\end{align}
where the uncertainties on $m_{\tilde{\mu}_1}$, $m_{\tilde{\mu}_2}$, and
$m_{\tilde\chi^0_1}$ are taken to be $0.2\GeV$, $0.2\GeV$, and $0.1\GeV$, respectively~\cite{Endo:2013xka}.
At the benchmark point of our choice, the uncertainty in the reconstructed value of $a_\mu^{(\tilde{B})}$ is dominated by the one related to the smuon left-right mixing parameter $m_{\tilde{\mu}{LR}}^2$.
It is concluded that the stau study at the ILC can help reconstructing $a_\mu^{(\tilde{B})}$ at the $1\%$ accuracy. Improvements of the measurements of the stau parameters can reduce the uncertainty, which may become possible with an ILC operation with higher luminosity or larger beam energy.

\section{Summary}\label{sec:summary}

The Fermilab experiment result confirmed the long-standing muon $g-2$ anomaly.
If the muon $g-2$ anomaly mainly comes from the Bino contribution in SUSY models, $\Damu\simeq \amu[SUSY]\simeq\amubino$, and if all the sleptons and the lightest neutralino are within the kinematical reach, the ILC will be able to measure the MSSM parameters which are necessary to estimate $\amubino$.
In this work,
we have investigated the measurements of the stau masses and mixing at the ILC, which are essential ingredients to determine the $\amubino$. We have also discussed their implication for the $\amubino$ reconstruction.
We have shown that, at our benchmark point, the SUSY contribution to the muon $g-2$ can be reconstructed with the accuracy of $\sim 1\%$ at the ILC with $\sqrt{s} = 500$~GeV and an integrated luminosity ${\cal L}=1.6 \iab$.

\section*{Acknowledgments}

We would like to thank D.~Jeans for valuable comments,
and the LCC generator working group and the ILD software working group for providing the simulation and reconstruction tools and producing the Monte Carlo samples used in this study.
This work has benefited from computing services provided by the ILC Virtual Organization, supported by the national resource providers of the EGI Federation and the Open Science GRID.
This work is supported in part by the Japan Society for the Promotion of Science (JSPS)  Grant-in-Aid for Scientific Research on Innovative Areas
(Nos.\,21H00086 [ME], 19H05810 [KH], 19H05802 [KH], and 16H06490 [TM]),
Scientific Research B (Nos.\,21H01086 [ME] and 20H01897 [KH]),
Scientific Research C (No.\,18K03608 [TM]),
and
Early-Career Scientists (Nos.\,16K17681 [ME] and 19K14706~[TK]).
The work is also supported by the JSPS Core-to-Core Program,
No.\,JPJSCCA20200002 [TK] and by World Premier International Research Center Initiative (WPI Initiative), MEXT, Japan.

\bibliographystyle{utphys28mod}
\bibliography{ref}
\end{document}